\documentclass[fleqn,usenatbib]{mnras}
\usepackage{amsmath,amssymb,graphicx}
\usepackage{multirow}
\usepackage{xspace}

\usepackage{hyperref}
\hypersetup{colorlinks=true,linkcolor=black,citecolor=black,urlcolor=black}

\usepackage{enumitem}

\newcommand{\hu}{\mathrm{km/s/Mpc}} 
\newcommand{\p}{{[P]}}
\newcommand{\oh}{{[\mathrm{O/H}]}}
\newcommand{\ngc}{NGC~4258\xspace}
\newcommand{\sn}{SN~Ia\xspace}
\newcommand{\sne}{SNe~Ia\xspace}
\newcommand{\MHW}{M_H^W}
\newcommand{\mpc}{\mathrm{Mpc}}
\newcommand{\sy}{\scriptstyle}

\newcommand{\td}{..}

\title[Prior sensitivity of the local distance ladder]{Physically-motivated priors in the local distance ladder significantly reduce the Hubble tension}

\author[M. H\"og\aa s \& E. M\"ortsell]{
Marcus H\"og\aa s$^{1,2}$\thanks{E-mail: marcus.hogas@fysik.su.se}
and Edvard M\"ortsell$^{1}$\thanks{E-mail: edvard@fysik.su.se}\\
$^{1}$Oskar Klein Centre, Department of Physics, Stockholm University,
Albanova University Center, 106 91 Stockholm, Sweden\\
$^{2}$Division of Mechanical Engineering, School of Information and Engineering, Dalarna University, Sweden
}

\date{Accepted XXX. Received YYY; in original form ZZZ}

\pubyear{2025}

\begin{document}

\label{firstpage}
\pagerange{\pageref{firstpage}--\pageref{lastpage}}

\maketitle

\begin{abstract}
Determinations of the Hubble constant based on the local distance ladder remain in significant tension with early-Universe inferences from the cosmic microwave background. While this tension is often discussed in terms of new physics or unmodeled systematics, the role of the assumed priors on the model parameters has received comparatively little attention.
Recently, Desmond et al. (2025) pointed out that the commonly adopted flat prior on distance moduli upweights smaller distances and systematically favors high inferred values of the Hubble constant.
Motivated by this observation, we perform a comprehensive Bayesian recalibration of the distance ladder, applying physically motivated priors uniformly to all distances, including the Milky Way Cepheids, which are incorporated directly into the joint fit. Together with a conservative treatment of the \emph{Gaia} EDR3 residual parallax offset, the Hubble constant shifts from $H_0 = 73.0 \pm 1.0 \, \mathrm{km/s/Mpc}$ to $H_0 = 70.6 \pm 1.0 \, \mathrm{km/s/Mpc}$, reducing the Hubble tension from $5 \, \sigma$ to $2 \, \sigma$.
Our results show that the assumed priors---often treated as innocuous defaults---may play a central role in the Hubble tension.
Because all local distance ladders rely on the calibration of distances, similar prior-driven effects are expected to arise across distance-ladder methods.
\end{abstract}

\begin{keywords}
cosmology: distance scale -- methods: statistical -- cosmology: cosmological parameters -- supernovae: general -- stars: variables: Cepheids
\end{keywords}

\section{Introduction}
\label{sec:Introduction}
Local distance-ladder measurements of the Hubble constant, \(H_0\), is in significant tension with early-Universe inferences within the standard \(\Lambda\)CDM\footnote{\(\Lambda\) Cold Dark Matter.} cosmological model. The most precise local determination is provided by the Cepheid--Type~Ia supernova (\sn) distance ladder calibrated by the SH0ES\footnote{Supernovae and \(H_0\) for the Equation of State of dark energy.} team, which in its fourth iteration yields \(H_0 = 73.0 \pm 1.0 \, \hu\) \citep{Riess:2021jrx}. In contrast, analyses of cosmic microwave background (CMB) anisotropies measured by \emph{Planck} infer a lower value, \(H_0 = 67.4 \pm 0.5 \, \hu\) \citep{Planck2020}, corresponding to a discrepancy of \(5\,\sigma\). While this ``Hubble tension'' has motivated extensive investigations of new physics and potential systematic effects, the role of statistical assumptions---particularly prior choices in the distance-ladder inference---has received comparatively little attention.

In Bayesian inference, priors are unavoidable.\footnote{The frequentist calibration of the distance ladder is discussed in Sec.~\ref{sec:Frequentist}.} In practice, however, they are often treated as innocuous defaults. In the standard SH0ES analysis, the distance-ladder parameters are inferred under the implicit assumption of flat priors. In particular, SH0ES adopts a flat prior in the \emph{distance moduli}, \(\pi(\mu)=\mathrm{const}\), which corresponds to an inverse prior in \emph{distances}, \(\pi(D)\propto D^{-1}\).\footnote{See Appendix~\ref{sec:DistTransf} for a derivation.} These priors systematically upweights smaller distances. As emphasized by \citet{Desmond:2025ggt}, a homogeneous, volume-limited population of distance indicators instead motivates a prior \(\pi(D)\propto D^{2}\), reflecting the increase in accessible volume with distance. Relative to the flat-in-modulus assumption, this physically-motivated prior assigns greater weight to larger distances and therefore leads, through Hubble’s law, to a lower inferred value of \(H_0\).

In this work, we perform a comprehensive recalibration of the fourth-iteration SH0ES distance ladder, incorporating Milky Way (MW) Cepheids directly into a single joint fit rather than treating them as external constraints. This enables a uniform application of physically motivated priors to all distance indicators and allows the residual \emph{Gaia} parallax offset to be constrained jointly.

Our central result is that adopting physically motivated distance priors, together with a conservative treatment of the MW parallax offset, coherently shifts the inferred distance ladder. The posterior favors distances that are larger by \( 3 \, \%\) on average and a Hubble constant that is lower by \(2.1 \, \hu\) relative to the reference SH0ES calibration, yielding \(H_0 = 70.6 \pm 1.0 \, \hu\). This reduces the tension between the local distance ladder and the \emph{Planck} CMB inference from \(5\,\sigma\) to \(2\,\sigma\). Our results demonstrate that the choice of priors may play a central role in the calibration of cosmic distances and in the Hubble tension.

\paragraph*{Notation.}
Throughout, the Hubble constant \(H_0\) is given in units of $\hu$ and the speed of light \(c\) in units of km/s.

\section{Data}
\label{sec:Data}

Our analysis uses the fourth-iteration SH0ES dataset, which comprises 42 calibrator \sne in 37 Cepheid host galaxies and is anchored by three geometric distance calibrators: MW, the Large Magellanic Cloud (LMC), and \ngc \citep{Riess:2021jrx}. The Cepheid sample is further supplemented by observations in M31 and the Small Magellanic Cloud (SMC). The dataset combines high-precision Cepheid photometry obtained primarily with the Hubble Space Telescope Wide Field Camera~3 (HST~WFC3) with standardized \sn measurements from the Pantheon+ compilation \citep{Riess:2021jrx,Scolnic:2021amr,Brout:2022vxf}.

We adopt the SH0ES data as compiled in \citet{MarcusSH0ESGitHub,Hogas:2024qlt}, which reproduces the SH0ES dataset exactly but reorganizes it to facilitate alternative calibrations of the distance ladder. Consequently, any difference in the inferred value of \(H_0\) in the present work arises exclusively from changes in the modeling and statistical assumptions, rather than from modifications to the underlying data.

In the MW, Cepheid distances are obtained from \emph{Gaia} parallaxes \citep{Riess:2020fzl}. The distance to \ngc is determined from very long baseline interferometric observations of water megamasers orbiting the central supermassive black hole, providing a purely geometric measurement with percent-level precision \citep{Reid:2019}. The LMC distance is calibrated using detached eclipsing binaries composed of late-type stars, yielding a geometric distance determination at the \( 1\%\) level \citep{Pietrzy_ski_2019}.

\section{Method}
\label{sec:Method}
The Cepheid--\sn distance ladder, as featured by the SH0ES team, links geometric distance measurements in the nearby Universe to \sne observed in the Hubble flow. Although it is naturally described in terms of three conceptual rungs, all parameters in this work are inferred simultaneously from a single joint likelihood.

\subsection{The three rungs}
The \emph{first rung} consists of the geometric distance anchors: MW, LMC, and \ngc. Cepheids are described using a reddening-free (Wesenheit) magnitude, $m_H^W$, with the period--luminosity relation
\begin{equation}
    \label{eq:PLR}
    m_{H,i}^W = \mu_i + \MHW + b_W \p_i + Z_W \oh_i ,
\end{equation}
where $\mu_i$ is the distance modulus and $\MHW$ the fiducial Cepheid absolute magnitude. We follow the standard SH0ES conventions
\begin{equation}
    \p_i \equiv \log_{10}\!\left(\frac{P_i}{10\,\mathrm{days}}\right), 
    \qquad
    \oh_i \equiv \log_{10}\!\left[\frac{(\mathrm{O/H})_i}{(\mathrm{O/H})_\odot}\right],
\end{equation}
so that $\p=0$ corresponds to a period of 10 days and $\oh=0$ to solar metallicity.

Geometric distance information enters the calibration differently for the three anchors. For the LMC and \ngc, it is incorporated through external constraints on the corresponding distance moduli, summarizing the geometric distance measurements \citep{Reid:2019,Pietrzy_ski_2019}. For MW Cepheids, geometric information is provided directly on a star-by-star basis through observed parallaxes \citep{Riess:2020fzl}. The observed parallax $\varpi_i$ (in mas) of Cepheid $i$ is related to its distance modulus via
\begin{equation}
\label{eq:ParallaxEq}
    \varpi_i = 10^{(10-\mu_i)/5} - zp ,
\end{equation}
where $zp$ denotes a residual parallax zero-point offset \citep{Lindegren_2021a,Lindegren_2021b}.

The \emph{second rung} consists of Cepheids in nearby ($\lesssim 80\,\mpc$) galaxies that host \sne. For a calibrator supernova $j$, the standardized rest-frame $B$-band apparent magnitude is
\begin{equation}
\label{eq:mB}
    m_{B,j} = M_B + \mu_j ,
\end{equation}
where $M_B$ is the fiducial \sn absolute magnitude.

The \emph{third rung} comprises supernovae in the Hubble flow. Using a low-redshift expansion of the distance--redshift relation, the standardized magnitude of the $k$th Hubble-flow supernova satisfies
\begin{align}
\label{eq:mBHF}
    &m_{B,k} - 5 \log_{10}\!\left[ c z_k \left\lbrace1 + \tfrac12 (1-q_0) z_k \right\rbrace \right] - 25 \nonumber \\
    &= M_B - 5 \log_{10} H_0 ,
\end{align}
where $q_0$ is the deceleration parameter, fixed to the SH0ES fiducial value $q_0=-0.55$.

Together, these relations define a single global distance-ladder model linking geometric distances, Cepheids, and supernovae.

\subsection{Likelihood}
The Cepheid--\sn distance-ladder model can be decomposed into a linear Gaussian component and an additional parallax-based component for the MW Cepheids. The linear component includes all Cepheid and supernova observables entering Eqs.~\eqref{eq:PLR}, \eqref{eq:mB}, and \eqref{eq:mBHF}, excluding the MW parallaxes, and can be written compactly as
\begin{equation}
    \mathbf{y} = L \mathbf{q},
\end{equation}
where $\mathbf{y}$ collects the observed magnitudes, $\mathbf{q}$ denotes the corresponding model parameters, and $L$ is the design matrix. The explicit forms of $\mathbf{y}$, $L$, and the covariance matrix $C$ are given in Appendix~\ref{app:SH0ESModel}.

Assuming Gaussian uncertainties, the likelihood of the linear component is
\begin{equation}
\label{eq:LikeLinear}
    \ln \mathcal{L}_{\rm lin}(\mathbf{q}) =
    -\frac12 \big( \mathbf{y} - L \mathbf{q} \big)^{T}
    C^{-1}
    \big( \mathbf{y} - L \mathbf{q} \big) .
\end{equation}
The MW Cepheids enter the same joint likelihood through an additional, non-linear term based on their observed \emph{Gaia} parallaxes. Denoting the model predictions implied by Eqs.~\eqref{eq:PLR} and \eqref{eq:ParallaxEq} by $m^{W,{\rm th}}_{H,i}$ and $\varpi_i^{\rm th}$, respectively, the likelihood contribution from MW Cepheid $i$ is
\begin{equation}
\label{eq:MWlike}
    \ln \mathcal{L}_{{\rm MW},i} =
    -\frac12 \left[
    \frac{(m^W_{H,i} - m^{W,{\rm th}}_{H,i})^2}{\sigma_{m,i}^2 + \sigma_{\rm int}^2}
    + \frac{(\varpi_i - \varpi_i^{\rm th})^2}{\sigma_{\varpi,i}^2}
    \right],
\end{equation}
where $\sigma_{m,i}$ and $\sigma_{\varpi,i}$ are the reported observational uncertainties. An intrinsic scatter of $\sigma_{\rm int}=0.07\,\mathrm{mag}$ is added in quadrature to the photometric uncertainty, consistent with the extragalactic Cepheids \citep{Riess:2021jrx}. The full MW log-likelihood is obtained by summing over the 66 MW Cepheids retained in the analysis,\footnote{Following the SH0ES team, we exclude two outliers.}
\begin{equation}
    \ln \mathcal{L}_{\rm MW} = \sum_{i=1}^{66} \ln \mathcal{L}_{{\rm MW},i}.
\end{equation}
The total log-likelihood is then
\begin{equation}
\label{eq:LikeTotal}
    \ln \mathcal{L} =
    \ln \mathcal{L}_{\rm lin} + \ln \mathcal{L}_{\rm MW},
\end{equation}
which is used throughout this work for statistical inference.

\subsection{Posterior}
By Bayes’ theorem, the posterior probability distribution of the model parameters is
\begin{equation}
    \ln \mathcal{P}(\theta) = \ln \mathcal{L}(\theta) + \ln \pi(\theta),
\end{equation}
where $\pi(\theta)$ denotes the prior distribution and $\theta$ collectively labels the model parameters.

The full parameter set comprises the distance moduli to the 37 \sn host galaxies; the anchor and non-\sn host distance moduli $(\mu_{\mathrm{N4258}}, \mu_{\mathrm{LMC}}, \mu_{\mathrm{M31}})$,\footnote{In the fourth-iteration SH0ES calibration, the distances to the LMC and the SMC are assumed to coincide.} the Cepheid PLR parameters $(M_H^W, b_W, Z_W)$; the \sn fiducial absolute magnitude $M_B$; the ground-to-HST photometric zero-point offset $\Delta \mathrm{zp}$; the Hubble constant parameterized by $5\log_{10}H_0$; the 66 MW Cepheid distance moduli $\{\mu_i\}$; and the residual parallax offset $zp$. In total, 113 model parameters.

\subsection{Sampling}
We sample the posterior using the \texttt{emcee} affine-invariant ensemble MCMC sampler \citep{GoodmanWeare,foreman2013emcee}. Convergence is ensured by requiring the total chain length to exceed \(100\) times the estimated autocorrelation time \(\tau\), supplemented by visual inspection of the chains. The initial \(2\tau\) samples are discarded as burn-in, and the number of walkers is set to twice the number of model parameters.

\subsection{Reference calibration (SH0ES-ref)}
\label{sec:SH0ESref}
As a reference calibration, we closely mimic the standard SH0ES analysis. This is, adopting flat priors on all model parameters, except for the residual parallax offset and the ground-to-HST photometric zero-point offset. For the former, we impose a Gaussian prior
\begin{equation}
\label{eq:zpPriorSH0ES}
    \pi(zp) = \mathcal{N}(0,\,0.01\,\mathrm{mas}),
\end{equation}
following \citet{Riess:2020fzl}, while for the ground-based photometric offset we adopt
\begin{equation}
    \pi(\Delta \mathrm{zp}) = \mathcal{N}(0,\,0.1\,\mathrm{mag}),
\end{equation}
following \citet{Riess:2021jrx}. With these choices, we obtain
\begin{equation}
\label{eq:H0ref}
    H_0 = 72.7 \pm 1.0 \, \hu \quad (\text{SH0ES-ref}).
\end{equation}
We refer to this setup as ``SH0ES-ref'' and adopt it as the reference for all subsequent calibrations. The resulting value of \(H_0\) is close to, but not identical with, the standard SH0ES result \(H_0 = 73.0 \pm 1.0 \, \hu\) \citep{Riess:2021jrx}. Since the underlying data are identical, the small offset (\(-0.3\,\hu\)) arises solely from methodological differences, specifically from incorporating the MW Cepheids directly into the joint distance-ladder fit rather than imposing them as external constraints on the fiducial Cepheid magnitude \(\MHW\).
As a consistency check, we also verified that our framework exactly reproduces the standard SH0ES result when the MW Cepheids are treated as external constraints.

\section{Results}
\label{sec:Results}
All results presented in this section are obtained by adopting the same likelihood, data vectors, covariance matrices, and model parametrization as in the reference calibration (SH0ES-ref) described in Sec.~\ref{sec:SH0ESref}.
The distance-ladder model is therefore held fixed throughout, and differences in the inferred value of \(H_0\) arise solely from changes in the assumed priors. 
We focus in particular on priors that are physically motivated, rather than chosen for practical convenience.

\subsection{Impact of physically motivated priors on $H_0$}
\label{sec:PhysPrior}
The reference calibration (SH0ES-ref) adopts flat priors on all distance moduli, mirroring the standard SH0ES implementation \citep{Riess:2021jrx}. A flat prior in distance modulus, however, is not neutral: since $\mu \propto \ln D$, $\pi(\mu)=\mathrm{const}$ corresponds to a distance prior $\pi(D)\propto D^{-1}$, which lacks a clear physical motivation. For a homogeneous, volume-limited population of distance indicators, the number of objects increases with distance squared, implying instead $\pi(D)\propto D^{2}$ \citep{Desmond:2025ggt}. Expressed in terms of distance modulus, this prior takes the form
\begin{equation}
\label{eq:muPrior}
    \ln \pi(\mu) = 0.6 \ln 10 \, \mu,
\end{equation}
up to an additive normalization constant (see App.~\ref{sec:DistTransf}). Relative to a flat prior in $\mu$, Eq.~\eqref{eq:muPrior} encodes the geometrical volume effect by assigning increasing weight to larger distances.

Here, we impose the prior Eq.~\eqref{eq:muPrior} uniformly on all distance moduli, including the MW Cepheids. This is enabled by our joint calibration framework, in which MW and extragalactic Cepheids are fitted simultaneously (Sec.~\ref{sec:Method}). In addition, we replace the Gaussian prior on the residual parallax offset, Eq.~\eqref{eq:zpPriorSH0ES}, with a flat prior,
\begin{equation}
\label{eq:zpPrior}
    \ln \pi(zp) = \mathrm{const},
\end{equation}
reflecting the remaining uncertainty in the \emph{Gaia} EDR3 parallax corrections \citep{Lindegren_2021a,Lindegren_2021b}. This conservative choice allows $zp$ to be constrained directly by the distance ladder itself (see \citealt{Hogas:2024qlt} for a detailed discussion).

\begin{figure}
    \centering
    \includegraphics[width=1\linewidth]{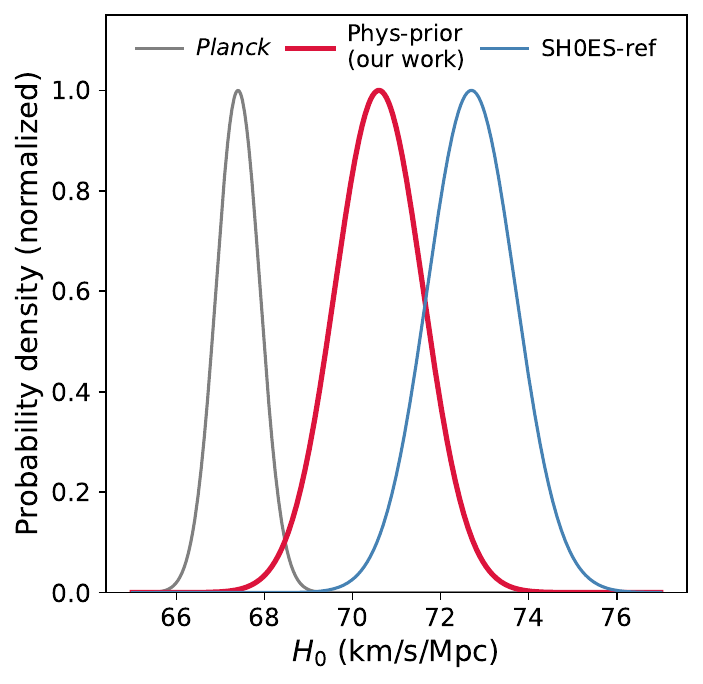}
    \caption{Normalized posterior distributions of the Hubble constant inferred under different prior assumptions. \emph{SH0ES-ref} corresponds to the reference calibration, yielding $H_0 = 72.7 \pm 1.0 \, \hu$. \emph{Phys-prior} shows the result obtained with the physically motivated distance priors and a flat prior on the residual \emph{Gaia} parallax offset, giving $H_0 = 70.6 \pm 1.0 \, \hu$. The \emph{Planck} curve indicates the CMB-based constraint, $H_0 = 67.4 \pm 0.5 \, \hu$. Under physically motivated priors, the tension with \emph{Planck} is reduced to $2\,\sigma$.}
    \label{fig:H0posteriors}
\end{figure}

With these prior choices, we obtain
\begin{equation}
\label{eq:H0PhysPrior}
    H_0 = 70.6 \pm 1.0 \, \hu .
\end{equation}
Relative to SH0ES-ref, this corresponds to a downward shift of $2.1 \, \hu$ in the Hubble constant, reducing the tension with the \emph{Planck} CMB constraint from $5\,\sigma$ to $2\,\sigma$ (Fig.~\ref{fig:H0posteriors}). Of the total shift, $-1.7 \, \hu$ is driven by the physically motivated distance prior Eq.~\eqref{eq:muPrior}, with the remainder arising from the flat prior on the parallax offset.

\begin{figure}
    \centering
    \includegraphics[width=1\linewidth]{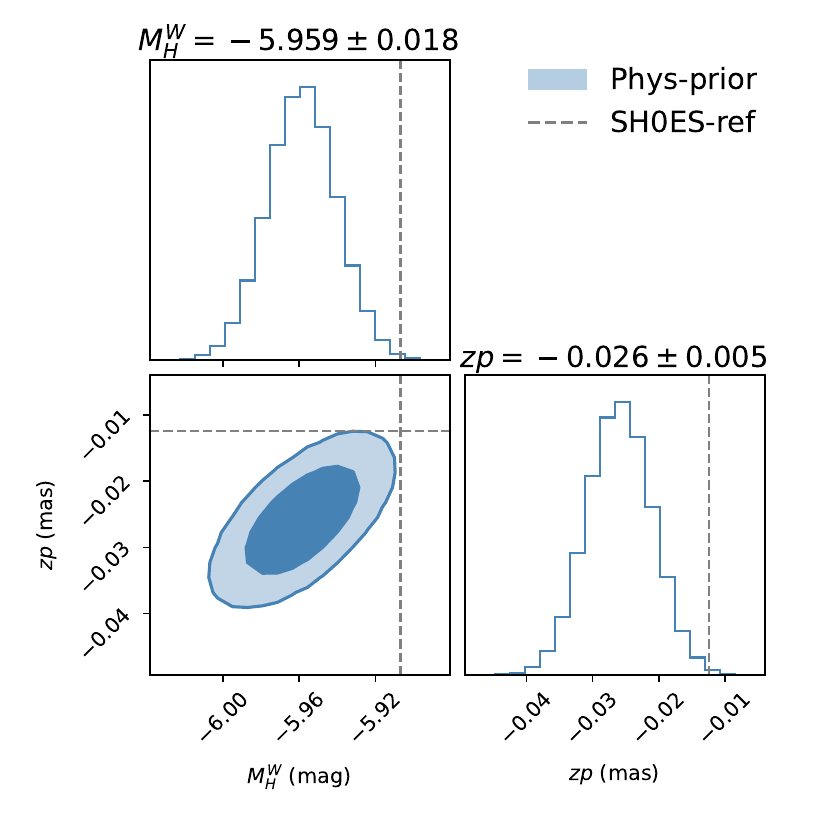}
    \caption{Marginalized posterior distributions in the $(\MHW, zp)$ plane for the Phys-prior (blue) and SH0ES-ref (dashed) calibrations. In SH0ES-ref, a Gaussian prior is imposed on $zp$, whereas in the Phys-prior case a flat prior is adopted. Allowing the data to constrain $zp$ directly shifts the posterior toward more negative values of $zp$ and brighter $\MHW$, contributing to the lower inferred value of $H_0$.}
    \label{fig:2Dmarg_MHW_zp}
\end{figure}

The role of the parallax-offset prior is illustrated in Fig.~\ref{fig:2Dmarg_MHW_zp}, which shows the marginalized posterior in the $(\MHW, zp)$ plane. When the residual parallax offset is constrained by the distance ladder itself, rather than by the informative SH0ES prior centered on zero (Eq.~\eqref{eq:zpPriorSH0ES}), the posterior shifts toward more negative values of $zp$. This shift is accompanied by a correlated change in $\MHW$, propagating through the ladder and contributing to a reduction in the inferred value of $H_0$.

The resulting constraint, $zp = -26 \pm 5 \, \mu\mathrm{as}$, is consistent with independent external estimates, many of which favor negative residual parallax offsets with typical values between $0$ and $-20 \, \mu\mathrm{as}$ \citep{Bhardwaj:2021,Fabricius:2021,Huang:2021,Zinn:2021}, while others allow for slightly positive values with substantially larger uncertainties \citep{Stassun:2021}.

\begin{figure*}
    \centering
    \includegraphics[width=1\linewidth]{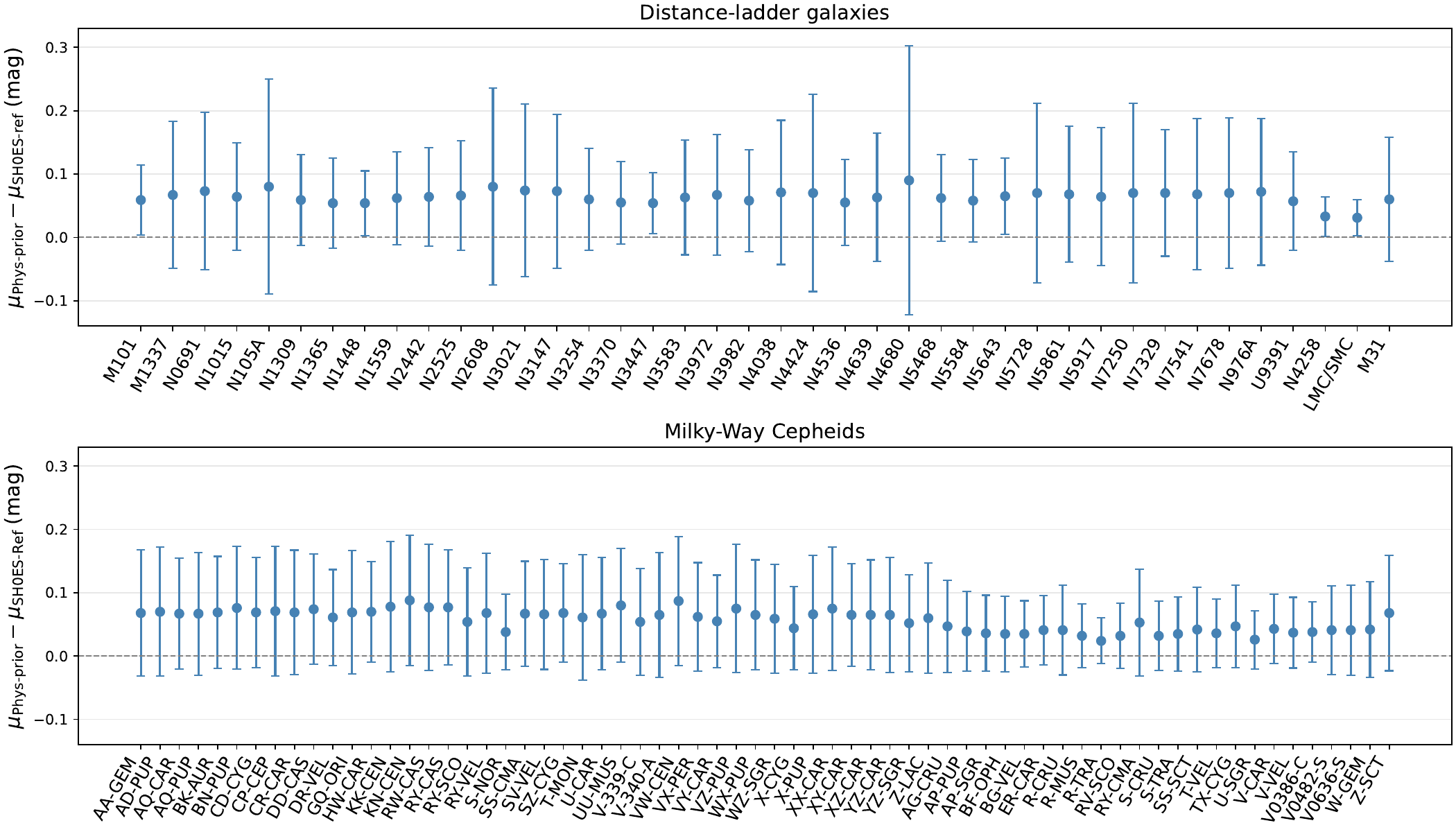}
    \caption{Shifts in inferred distance moduli relative to the reference calibration, $\mu_{\rm Phys\text{-}prior} - \mu_{\rm SH0ES\text{-}ref}$. \textit{Top:} distance-ladder galaxies. \textit{Bottom:} MW Cepheids. Error bars show posterior uncertainties from SH0ES-ref. The physically motivated distance prior produces coherent positive shifts, larger for more weakly constrained distances. The mean shifts are $0.066\,\mathrm{mag}$ for \sn hosts and $0.057\,\mathrm{mag}$ for MW Cepheids, corresponding to average distance increases of $3.0\%$ and $2.6\%$, respectively, which drive the associated change in $H_0$.}
    \label{fig:Delta_mu}
\end{figure*}

\subsection{Distance-modulus shifts}
Figure~\ref{fig:Delta_mu} shows the shifts in inferred distance moduli,
$\mu_{\rm Phys\text{-}prior} - \mu_{\rm SH0ES\text{-}ref}$, for the distance-ladder galaxies (top) and the individual MW Cepheids (bottom). As expected, the physically motivated prior in Eq.~\eqref{eq:muPrior} produces positive shifts in all distances. Averaged over the sample, the \sn host galaxies exhibit a mean increase of $0.066\,\mathrm{mag}$, corresponding to a $3.0\%$ increase in distance, while the MW Cepheids show a mean shift of $0.057\,\mathrm{mag}$, or $2.6\%$ in distance.

The magnitude of the shift correlates strongly with the statistical uncertainty of each object: distances that are more weakly constrained by the likelihood increase the most. This behavior reflects the structure of the prior in Eq.~\eqref{eq:muPrior}, whose influence is strongest where the data are least informative.

These coherent distance increases propagate directly into the inferred value of the Hubble constant. From Hubble’s law, $cz = H_0 D$, a fractional increase in distance implies an equal fractional decrease in $H_0$. The $3\%$ mean distance shifts therefore correspond to a comparable reduction in the Hubble constant, yielding $H_0 \simeq 70.5\,\hu$ starting from the SH0ES-ref value $H_0 = 72.7\,\hu$, in excellent agreement with the full joint inference $H_0 = 70.6 \pm 1.0\,\hu$.

\begin{figure}
    \centering
    \includegraphics[width=1\linewidth]{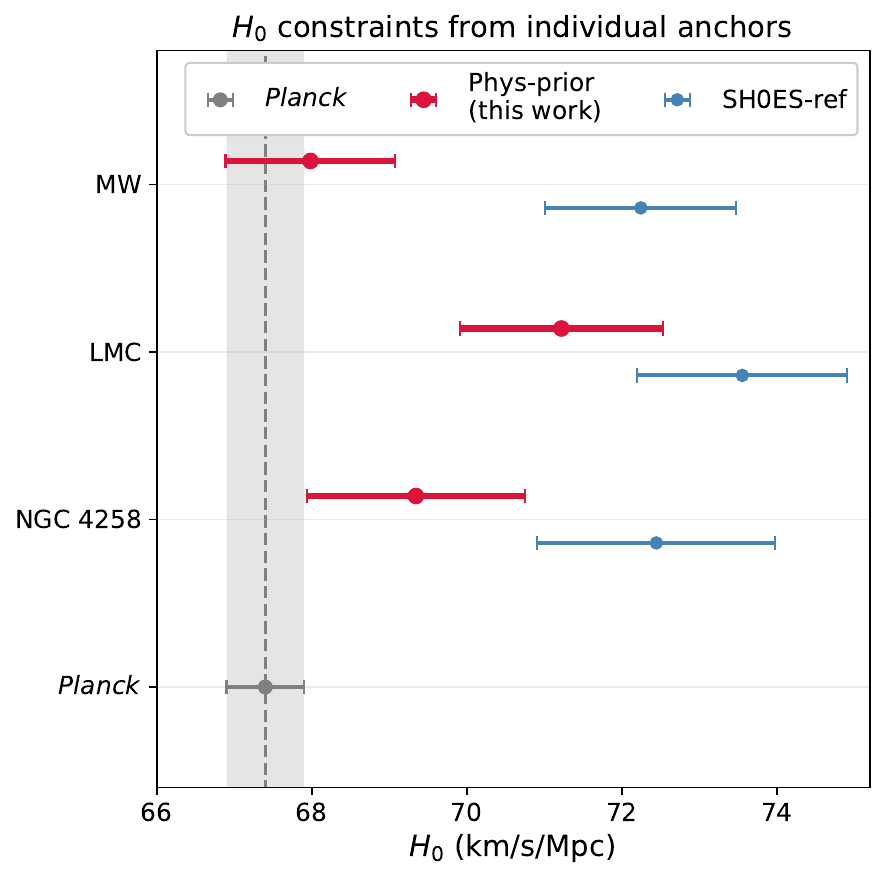}
    \caption{Constraints on the Hubble constant inferred from the local distance ladder when anchored to individual geometric calibrators. Shown are marginalized $H_0$ constraints obtained using the MW, the LMC, and \ngc\ as anchors for the SH0ES-ref calibration (blue) and the calibration with physically motivated priors (red). The horizontal gray band indicates the \emph{Planck} CMB constraint. Physically motivated distance priors shift $H_0$ downward for all anchors while preserving their mutual consistency, thereby reducing the tension with the CMB.}
    \label{fig:H0_anch}
\end{figure}

\subsection{Anchor-by-anchor calibrations}
We also examine how the choice of priors affects the inferred $H_0$ when the distance ladder is anchored to single geometric calibrators. Figure~\ref{fig:H0_anch} shows the marginalized constraints obtained when using the MW, the LMC, or \ngc\ as the primary anchor, for both the reference calibration (SH0ES-ref) and the calibration with physically motivated priors (Phys-prior).

In the SH0ES-ref case, the anchor-by-anchor determinations of $H_0$ are mutually consistent within uncertainties and cluster around the combined value in Eq.~\eqref{eq:H0ref}. When adopting physically motivated distance priors, all anchor-based determinations shift coherently toward lower values of $H_0$. This demonstrates that the reduction in $H_0$ is not driven by any single distance calibrator but reflects a global rescaling of the distance ladder induced by the modified prior assumptions.

The shift is largest for the MW-anchored calibration, as expected, since it is affected both by the distance-modulus prior and by the treatment of the residual parallax offset. The LMC- and \ngc-anchored calibrations exhibit shifts in the same direction but with smaller amplitudes. Importantly, the relative ordering of the anchor-based $H_0$ values and their uncertainties remain unchanged between the two calibrations, indicating that the physically motivated priors do not introduce additional internal tension between anchors. For reference, the \emph{Planck} CMB constraint is also shown in Fig.~\ref{fig:H0_anch}, illustrating the substantial reduction of the Hubble tension under the Phys-prior calibration.

\section{Discussion}
\label{sec:Discussion}
\subsection{Motivating the distance prior}
In the Phys-prior calibration, we adopt a volume-weighted distance prior, Eq.~\eqref{eq:muPrior}, corresponding to a homogeneous, volume-limited population of distance indicators. Here we assess how well these assumptions are satisfied by the Cepheid--\sn distance ladder and argue that this prior provides a reasonable baseline.

The SH0ES sample of \sn host galaxies is commonly described as a complete set of suitable \sne discovered at redshifts $z \leq 0.01$ \citep{Riess:2021jrx} and any residual observational selection effects are explicitly corrected for in the supernova analysis (see Sec.~\ref{sec:BBC}), making the assumption of a volume-limited sample a reasonable working approximation.

For the MW Cepheids, which are intrinsically luminous, modern facilities such as \emph{Gaia} and HST are sensitive to distances well beyond those spanned by the MW sample used in the calibration. The MW Cepheid sample can therefore be assumed to be close to complete.

The assumption of homogeneity need only hold over the distance range where the posterior has significant support, which is set by the distance uncertainty of each Cepheid. To estimate the expected level of inhomogeneity at this scale, we model the Galactic stellar density as a standard double-exponential disk with radial scale lengths $R_{d,\mathrm{thin}} \simeq 2.5\,\mathrm{kpc}$ and $R_{d,\mathrm{thick}} \simeq 3.4\,\mathrm{kpc}$ \citep{McMillan:2016jtx}. An \emph{upper limit} for the density variation can be estimated by assuming that the Cepheids lie along the radial direction toward the Galactic center. In this case, the fractional density variation across the region probed by the posterior is $\Delta\rho/\rho \simeq \sigma_R / R_d$, where $\sigma_R$ is the distance uncertainty.
With an average $\sigma_R \simeq 0.06\,\mathrm{kpc}$, this implies upper-limit density variations of $\lesssim 3\%$ for the thin disk and $\lesssim 2\%$ for the thick disk.
Thus, the MW Cepheid population can be well approximated as locally homogeneous over the distance range effectively probed by the likelihood, supporting the use of a volume-weighted distance prior, Eq.~\eqref{eq:muPrior}.

\subsection{Distance priors and the Hubble-flow \sne}
Contrary to \citet{Desmond:2025ggt}, we apply the volume-weighted distance prior, Eq.~\eqref{eq:muPrior}, only to distances that enter explicitly as fitted parameters. It is not applied to the Hubble-flow \sne, for which distances are only derived quantities.
For the Hubble flow \sne, the derived distances are inversely proportional to the (fitted) Hubble constant and thus upweighting large distances in the Hubble flow is equivalent to imposing a prior that favors smaller values of \(H_0\). Such a choice would amount to a prior on \(H_0\) itself rather than on the distances. The impact of different prior assumptions on \(H_0\) is examined separately in Appendix~\ref{sec:Robustness}.

\subsection{Frequentist analysis}
\label{sec:Frequentist}
The SH0ES team presents both a frequentist maximum-likelihood calibration of the distance ladder and a Bayesian inference of the same model \citep{Riess:2021jrx}. Owing to the linear structure of their model, the maximum-likelihood solution can be obtained analytically. When flat priors are adopted on the model parameters, the Bayesian inference yields identical results.
This coincidence between the frequentist and Bayesian results is not fundamental, but hinges on the assumption of flat priors. For linear Gaussian models, flat Bayesian priors correspond exactly to the maximum-likelihood solution. The agreement between the two approaches therefore reflects a particular prior choice.
More generally, prior assumptions in the Bayesian framework correspond to penalty terms in the frequentist framework. Physically motivated assumptions about distances—such as those explored in this work—can thus be implemented consistently in either statistical formulation. The resulting change in the inferred value of $H_0$ is driven by these assumptions themselves, not by the choice of Bayesian or frequentist methodology.

\subsection{Selection effects and the BBC framework}
\label{sec:BBC}
The BEAMS with Bias Corrections (BBC) method \citep{Popovic2021} corrects \sn distance moduli for observational selection effects and standardization biases using detailed Monte--Carlo simulations of the surveys. Synthetic supernovae are propagated through the same detection and analysis pipeline as the data, allowing biases in the fitted distance moduli to be quantified by comparison with the known input values. The resulting bias,
\begin{equation}
    \delta\mu_{\rm bias}(z,X_1,C,M_\ast,\alpha,\beta)
    = \mu_{\rm fit} - \mu_{\rm true},
\end{equation}
is tabulated on a multidimensional grid and interpolated to assign a correction to each observed \sn as a function of supernova and host-galaxy properties.

These corrections are applied prior to the cosmological inference and ensure that the observed \sn sample provides an unbiased estimate of relative distances by accounting for correlations between supernova properties, host environment, and survey selection. 
The BBC framework thus operates at the level of the \sn standardization and is not designed to compensate for assumptions about distance priors in the calibration of the local distance ladder, which enter at a different stage of the inference. We therefore argue that the suggestion in \citet{Desmond:2025ggt} that the BBC method may compensate for the use of a flat prior in distance modulus does not apply.




\section{Conclusions}
\label{sec:Conclusions}
We have examined how prior assumptions on model parameters—most importantly on distances—affect the calibration of the local Cepheid--\sn distance ladder and the inferred value of the Hubble constant. A coherent prior-driven shift of all distances by only a few percent propagates directly into a comparable shift in $H_0$.

Adopting physically motivated priors—a volume-weighted distance prior $\pi(D)\propto D^2$ and a conservative flat prior on the \emph{Gaia} residual parallax offset—we find
\begin{equation}
    H_0 = 70.6 \pm 1.0 \, \hu,
\end{equation}
compared to $H_0 = 73.0 \pm 1.0 \, \hu$ in the standard SH0ES calibration \citep{Riess:2021jrx}. This reduces the tension with the \emph{Planck} CMB inference from $5\,\sigma$ to $2\,\sigma$ and demonstrates that the Hubble tension depends sensitively on prior choices that are often treated as innocuous defaults.

While our analysis focuses on the Cepheid--\sn ladder, the underlying considerations are generic to all distance-ladder methods, including Tip of the Red Giant Branch (TRGB), Mira variable stars, surface brightness fluctuations (SBF), J-region asymptotic giant branch stars (JAGB), as well as approaches based on the Fundamental Plane and the Tully--Fisher relation. In all such cases, distances enter as model parameters and must therefore be assigned priors whose influence may be significant.


\section*{Acknowledgements}
MH and EM acknowledge support from the Swedish Research Council under Dnr VR 2024-03927.

OpenAI's ChatGPT has been used to draft portions of the manuscript.

\section*{Data availability}
The fourth-iteration SH0ES data is publicly available on GitHub: \citet{SH0ESData} or alternatively \citet{MarcusSH0ESGitHub}.

\appendix
\section{The SH0ES-ref Model}
\label{app:SH0ESModel}
In this appendix we present in detail the linear formulation of the Cepheid--\sn distance ladder used in the SH0ES-ref calibration (Sec.~\ref{sec:Method}).
The model can be written as
\begin{equation}
    \mathbf{y} = L \mathbf{q},
\end{equation}
where $\mathbf{y}$ is the data vector, $\mathbf{q}$ is the vector of model parameters, and $L$ is the design matrix.
The starting point is the fourth-iteration SH0ES data, compiled and publicly available on GitHub: \citet{MarcusSH0ESGitHub}.
Our data vector $\textbf{y}$, design matrix $L$, and covariance matrix $C$ are taken directly from \citet{MarcusSH0ESGitHub}, with two modifications:
\begin{enumerate}[leftmargin=*, labelsep=0.5em]
    \item The external constraints due to the MW Cepheids are removed, since in the SH0ES-ref calibration the MW Cepheids are incorporated directly through an explicit likelihood term (Eq.~\ref{eq:MWlike}).

    \item We treat the ground-to-HST zero-point offset, $\Delta\mathrm{zp}=0\pm0.1\,\mathrm{mag}$, as a Gaussian prior rather than as an external constraint, $\Delta\mathrm{zp} = \mathcal{N}(0, \, 0.1 \, \mathrm{mag})$.
\end{enumerate}
Apart from these adjustments, our $\textbf{y}$, $L$, and $C$ are identical to \citet{MarcusSH0ESGitHub}.
The data vector $\mathbf{y}$ collects all Cepheid and supernova observables entering the three rungs, together with a small number of external constraints. Schematically, it can be written as in Eq.~\eqref{eq:y_R22} and contains the following entries:
\begin{itemize}[leftmargin=*, labelsep=0.5em]
  \item Wesenheit magnitudes $m_H^W$ of Cepheids observed in SN~Ia host galaxies,
  
  \item Wesenheit magnitudes of Cepheids in anchor galaxies and non-\sne host galaxies (\ngc, M31, LMC, and SMC),

  \item standardized apparent magnitudes of \sne in Cepheid-calibrated host galaxies

  \item standardized apparent magnitudes of \sne in the Hubble flow, combined with the low-redshift expansion of the distance--redshift relation, and

  \item the geometric anchor distances to LMC and \ngc as external constraints.
\end{itemize}
\begin{figure*}
    \centering
    
        \begin{equation}
        \label{eq:y_R22}
        \mathbf{y} = 
        \begin{array}{ll}
        \left(
        \begin{array}[c]{c}
        
        m^W_{H,\mathrm{M101}} \\
    
        : \\
    
        m^W_{H,\mathrm{U9391}} \\
        
        \hline
        
        m^W_{H,\textrm{N4258}} \\
        
        m^W_{H,\textrm{M31}} \\
        
        m^W_{H,\textrm{LMC,GRND}} \\
    
        m^W_{H,\textrm{LMC,HST}} \\
        
        m^W_{H,\textrm{SMC}} \\
        
        \hline
        
        m_{B} \\
    
        \hline
        
        m_{B} - 5 \log_{10} \left[ c z \{ ... \} \right] -25 \\
        
        \hline
        
        \mu_\mathrm{N4258}^\mathrm{anch} \\
        
        \mu_\mathrm{LMC}^\mathrm{anch}
        
        \end{array} \right)
        
        &
        
        \begin{array}[c]{@{}l@{\,}l}
        
        \left.
        \begin{array}{c} \vphantom{m^W_{H,\mathrm{hosts}}} \\
        \vphantom{:} \\
        \vphantom{m^W_{H,\mathrm{hosts}}}
        \end{array}
        \right\} & \text{2150 Cepheids in SNIa hosts} \\
        
        \left.
        \begin{array}{c} \vphantom{m^W_{H,\textrm{nh},j}} \\ 
        \vphantom{m^W_{H,\textrm{nh},j}} \\ 
        \vphantom{m^W_{H,\textrm{nh},j}} \\
        \vphantom{m^W_{H,\textrm{nh},j}} \\
        \vphantom{\textrm{\LARGE HELLO}}
        \end{array}
        \right\} & \text{(443 + 55 + 270 + 69 + 143) Cepheids in anchors or non-SNIa hosts\hspace{1.5in}} \\
        
        \left.
        \begin{array}{c}
        \vphantom{m_B^0}
        \end{array}
        \right\} & \text{77 Calibrator SN Ia magnitudes} \\
    
        \left.
        \begin{array}{c}
        \vphantom{.}
        \end{array}
        \right\} & \text{277 Hubble-flow SNe Ia} \\
        
        \left.
        \begin{array}{c}
        \vphantom{.} \\
        \vphantom{.}
        \end{array}
        \right\} & \text{2 External constraints}
        
        \end{array}
        \end{array}
    \end{equation}
\end{figure*}
The vector of linear model parameters, $\textbf{q}$, is shown in Eq.~\eqref{eq:q}.
\begin{figure}
    \centering
    
    \begin{equation}
    \label{eq:q}
    \mathbf{q} = 
        \left(
        \begin{array}{c}
            \mu_\mathrm{M101} \\
            : \\
            \mu_\mathrm{U9391} \\
            \hline
            \mu_\mathrm{N4258} \\
            \MHW \\
            \mu_\mathrm{LMC} \\
            \mu_\mathrm{M31} \\
            b_W \\
            M_B \\
            Z_W \\
            \Delta \mathrm{zp} \\
            5 \log_{10} H_0
        \end{array}
        \right) .
        \end{equation}

\end{figure}
The design matrix $L$ encodes the linear mapping between the parameters $\mathbf{q}$ and the observables $\mathbf{y}$. Each row of $L$ corresponds to one element of the data vector and selects the appropriate linear combination of model parameters entering the Cepheid PLR Eq.~\eqref{eq:PLR}, the \sn magnitude--distance relation Eq.~\eqref{eq:mB}, or the Hubble-flow relation Eq.~\eqref{eq:mBHF}.
The explicit block structure of $L$ is given in Eq.~\eqref{eq:L_R22}.
\begin{figure*}
    \centering

    \begin{equation}
        \label{eq:L_R22}
        L =
        \left(
        \begin{array}[c]{ccccccclclcc}
        
        1 & \td & 0 & 0 & 1 & 0 & 0 & [P]_\mathrm{M101} & 0 & \oh_\mathrm{M101} & 0 & 0 \\
        
        : & \rotatebox{45}{:} & : & : & : & : & :   & : & : & : & : & : \\  
        
        0 & \td & 1 & 0 & 1 & 0 & 0 & [P]_\mathrm{U9391} & 0 & \oh_\mathrm{U9391} & 0 & 0 \\
        
        \hline
        
        0 & \td & 0 & 1 & 1 & 0 & 0 & [P]_\mathrm{N4258} & 0 & \oh_{\textrm{N4258}} & 0 & 0 \\
        
        0 &\td & 0 & 0 & 1 & 0 & 1 & [P]_\mathrm{M31} & 0 & \oh_{\textrm{M31}}   & 0 & 0 \\
        
        0 & \td & 0 & 0 & 1 & 1 & 0 & [P]_\mathrm{LMC,GRND} & 0 & \oh_{\textrm{LMC,GRND}} & 1 & 0 \\

        0 & \td & 0 & 0 & 1 & 1 & 0 & [P]_\mathrm{LMC,HST} & 0 & \oh_{\textrm{LMC,HST}} & 0 & 0 \\

        0 & \td & 0 & 0 & 1 & 1 & 0 & [P]_\mathrm{SMC} & 0 & \oh_{\textrm{SMC}}    & 1 & 0 \\
        
        \hline
        
        1 & \td & 0 & 0 & 0 & 0 & 0 & 0 & 1 & 0 & 0 & 0 \\
        
        : & \rotatebox{45}{:} & : & : & : & : & : & : & : & : & : & : \\  
        
        0 & \td & 1 & 0 & 0 & 0 & 0 & 0 & 1 & 0 & 0 & 0 \\
        
        \hline

        0 & \td & 0 & 0 & 0 & 0 & 0 & 0 & 1 & 0 & 0 & -1 \\

        : & \rotatebox{45}{:} & : & : & : & : & : & : & : & : & : & : \\
        
        0 & \td & 0 & 0 & 0 & 0 & 0 & 0 & 1 & 0 & 0 & -1\\

        \hline

        0 & \td & 0 & 1 & 0 & 0 & 0 & 0 & 0 & 0 & 0 & 0 \\
        
        0 & \td & 0 & 0 & 0 & 1 & 0 & 0 & 0 & 0 & 0 & 0
        
        \end{array}
        \right)
    \end{equation}
\end{figure*}

The covariance matrix $C$ contains the full uncertainties of the data vector $\mathbf{y}$, including measurement uncertainties, intrinsic scatter terms, and correlated systematic contributions, such as metallicity-dependent correlations among the galaxies and the full covariance of the \sn sample.
The matrix has a block-diagonal structure reflecting the different components of the data vector. The explicit form of $C$ used in this work is provided in Eq.~\eqref{eq:C_R22}.

\begin{figure*}
    \centering

    \begin{equation}
    \label{eq:C_R22}
        C = \left(
        \begin{array}{cccccccccccccccc}
        
        \sy{\sigma_{\rm M101}^2}\!\!\!\! & \td & \sy{Z_{\textrm{cov}}} & \sy{Z_{\textrm{cov}}} & \sy{0} & \sy{0} & \sy{0} & \sy{0} & \sy{0} & \sy{0}  & \sy{0} & \sy{0} \\
        
        : & \rotatebox{45}{:} & : & : & : & : & : & : & : & : & : & : \\
        
        \sy{Z_{\textrm{cov}}} & \td & \sy{\sigma_{\rm U9391}^2}\!\!\!\! & \sy{Z_{\textrm{cov}}} & \sy{0} & \sy{0} & \sy{0} & \sy{0} &\sy{0} & \sy{0} & \sy{0} & \sy{0} \\
        
        \hline
        
        \sy{Z_{\textrm{cov}}} & \td & \sy{Z_{\textrm{cov}}} & \sy{\sigma_{{\rm N4258}}^2}\!\!\!\! & \sy{0} & \sy{0} & \sy{0} & \sy{0}  &\sy{0} & \sy{0} & \sy{0} & \sy{0} \\
        
        \sy{0} & \td & \sy{0} & \sy{0} & \sy{\sigma_{{\rm M31}}^2}\!\!\!\! & \sy{0} & \sy{0} & \sy{0}  &\sy{0} & \sy{0} & \sy{0} & \sy{0} \\

        \sy{0} & \td & \sy{0} & \sy{0} & \sy{0} &  \sy{\sigma_{{\rm LMC,GRND}}^2}\!\!\!\! & \sy{10^{-4}} & \sy{0} & \sy{0} & \sy{0} & \sy{0} & \sy{0} \\

        \sy{0} & \td & \sy{0} & \sy{0} & \sy{0} & \sy{10^{-4}} &  \sy{\sigma_{{\rm LMC,HST}}^2}\!\!\!\! & \sy{0} & \sy{0} & \sy{0} & \sy{0} & \sy{0} \\

        \sy{0} & \td & \sy{0} & \sy{0} & \sy{0} & \sy{0} & \sy{0} &  \sy{\sigma_{{\rm SMC}}^2}\!\!\!\! & \sy{0} & \sy{0} & \sy{0} & \sy{0} \\

        \hline
                
        \sy{0} & \td & \sy{0} & \sy{0} & \sy{0} & \sy{0} & \sy{0} & \sy{0} & \sy{\sigma^2_{\textrm{Cal SN}}}\!\!\!\! & \sy{{\rm SN}_{\textrm{cov}}} & \sy{0} & \sy{0} \\
        
        \hline

        \sy{0} & \td & \sy{0} & \sy{0} & \sy{0} & \sy{0} & \sy{0} & \sy{0} & \sy{{\rm SN}_{\textrm{cov}}} & \sy{\sigma^2_{\textsc{HF SN}}}\!\!\!\! & \sy{0} & \sy{0} \\
        
        \hline
        
        \sy{0} & \td & \sy{0} & \sy{0} & \sy{0} & \sy{0} & \sy{0} & \sy{0} & \sy{0} & \sy{0} & \sy{\sigma_{\mu,{\rm N4258}}^2}\!\!\!\! & \sy{0} \\
        
        \sy{0} & \td & \sy{0} & \sy{0} & \sy{0} & \sy{0} & \sy{0} & \sy{0} &  \sy{0} & \sy{0} & \sy{0} & \sy{\sigma_{\mu,{\rm LMC}}^2}
    \end{array}
    \right)
    \end{equation}
    
\end{figure*}

\section{Transformation of the distance prior}
\label{sec:DistTransf}
Here, we derive the prior on the distance modulus implied by a power-law prior on the distance.
Let the prior on the physical distance $D$ be
\begin{equation}
    \pi(D) \propto D^{k}.
\end{equation}
The distance modulus is related to the distance by
\begin{equation}
    \mu = 5 \log_{10} \left( \frac{D}{10 \, \mathrm{pc}} \right),
\end{equation}
which can be inverted to give
\begin{equation}
    D = 10^{(\mu+5)/5} \, \mathrm{pc}.
\end{equation}
Under a change of variables, the prior transforms as
\begin{equation}
    \pi(\mu) = \pi(D) \left| \frac{\mathrm{d}D}{\mathrm{d}\mu} \right|.
\end{equation}
Using $\mathrm{d}D/\mathrm{d}\mu = (\ln 10 / 5) D$, this yields
\begin{equation}
    \pi(\mu) \propto D^{k+1} \propto 10^{\frac{k+1}{5}\mu}.
\end{equation}
Equivalently, up to an additive normalization constant,
\begin{equation}
\label{eq:pi_mu_k}
    \ln \pi(\mu) = \frac{k+1}{5} \ln 10 \, \mu .
\end{equation}
For a homogeneous, volume-limited sample, $k=2$ which reduces to
\begin{equation}
    \ln \pi(\mu) = 0.6 \ln 10 \, \mu.
\end{equation}
That is, the distance-modulus prior adopted in Sec.~\ref{sec:Results}.

\section{Robustness to prior assumptions}
\label{sec:Robustness}
\begin{table*}
\centering
\caption{One-at-a-time prior variations around the reference calibration (SH0ES-ref). In each row we modify the prior on a single parameter (or set of parameters) while keeping everything else identical with the SH0ES-ref setup (Sec.~\ref{sec:Method}). We report the resulting $H_0$ constraint and its shift relative to SH0ES-ref. The first row is the SH0ES-ref calibration and the last row the Phys-prior calibration.}
\label{tab:PriorRobustness}
\begin{tabular}{l l l c c}
\hline
\textbf{Prior(s) modified} &
\textbf{SH0ES-ref prior} &
\textbf{Alternative prior} &
\textbf{$H_0$ [$\hu$]} &
\textbf{$\Delta H_0$ [$\hu$]}\\

\hline
(none) &
(as in Sec.~\ref{sec:Method}) &
(none) &
$72.7 \pm 1.0$ &
--- \\

\hline
All distance moduli (incl. MW) $\mu$ &
flat &
$\pi(D)\propto D^{2}$ &
$71.0 \pm 1.0$ &
$-1.7$ \\

\emph{Gaia} EDR3 residual parallax offset $zp$ &
$\pi(zp)=\mathcal{N}(0,\,0.01\,\mathrm{mas})$ &
flat &
$72.5 \pm 1.0$ &
$-0.2$ \\

The Hubble constant $H_0$ &
flat in $5\log_{10}H_0$ &
flat &
$72.7 \pm 1.0$ &
$\pm 0.0$ \\

All fiducial magnitudes ($\MHW$ and $M_B$) &
flat &
flat in luminosities &
$72.7 \pm 1.0$ &
$\pm 0.0$ \\

PLR slope $b_W$ &
flat &
$\pi(b_W) \propto |b_W|$ &
$72.7 \pm 1.0$ &
$\pm 0.0$ \\

PLR slope $b_W$ &
flat &
$\pi(b_W) \propto 1/|b_W|$ &
$72.7 \pm 1.0$ &
$\pm 0.0$ \\

Metallicity--luminosity slope $Z_W$ &
flat &
$\pi(Z_W) \propto |Z_W|$ &
$72.7 \pm 1.0$ &
$\pm 0.0$ \\

Metallicity--luminosity slope $Z_W$ &
flat &
$\pi(Z_W) \propto 1/|Z_W|$ &
$72.7 \pm 1.0$ &
$\pm 0.0$ \\

Ground-to-HST offset $\Delta\mathrm{zp}$ &
$\mathcal{N}(0,\,0.1\,\mathrm{mag})$ &
flat &
$72.7 \pm 1.0$ &
$\pm 0.0$ \\

\hline
All distance moduli + $zp$  &
(as in Sec.~\ref{sec:Method}) &
(as in Sec.~\ref{sec:Results}) &
$70.6 \pm 1.0$ &
$-2.1$ \\

\hline
\end{tabular}
\end{table*}

Table~\ref{tab:PriorRobustness} summarizes the impact of varying prior assumptions one set of parameters at a time around the reference calibration (SH0ES-ref). In each case, only the specified prior is changed, while all others are held fixed to their SH0ES-ref values. The table covers all model parameters, including those not modified in the baseline Phys-prior calibration. This exercise serves as a comprehensive diagnostic of the sensitivity of $H_0$ to individual prior choices, rather than as an attempt to motivate any particular choice of priors.

Two conclusions emerge. First, adopting the physically motivated distance prior $\pi(D)\propto D^2$ for all distance moduli, including the MW Cepheids, leads to a downward shift of $-1.7\,\hu$ relative to SH0ES-ref. Relaxing the Gaussian prior on the residual parallax offset in favor of a more conservative, flat, prior produces a smaller but non-negligible shift of $-0.2\,\hu$. When both of these physically motivated priors are applied simultaneously, the total shift amounts to $-2.1\,\hu$, yielding $H_0 = 70.6 \pm 1.0\,\hu$.

Second, the calibration of $H_0$ is insensitive to prior assumptions on the remaining model parameters. Changing the prior on the Hubble constant itself, switching between flat priors in magnitude and luminosity for the Cepheid and \sn fiducial magnitudes, or adopting scale-invariant or scale-weighted priors for the PLR slope $b_W$ and the metallicity--luminosity slope $Z_W$, leaves the inferred value of $H_0$ unchanged at the level of $< 0.1\,\hu$. Likewise, relaxing the Gaussian prior on the ground-to-HST photometric offset $\Delta\mathrm{zp}$ has a negligible effect. This demonstrates that the distance ladder is robust to reasonable variations in priors on these parameters and that the data dominate their inference.

Taken together, these results show that the dominant prior sensitivity of the distance ladder resides in the distances and, to a lesser degree, in the residual parallax offset.

\bibliographystyle{mnras}
\bibliography{bibliography}
\label{lastpage}
\end{document}